\documentclass[10pt]{article}
\def\be{\begin{equation}}
\def\ee{\end{equation}}
\begin{document}

\title{The noncommutative replica procedure}
\author{S.V.Kozyrev}

\maketitle

\begin{abstract}
The alternative to the replica procedure, which we call the
noncommutative replica procedure, is discussed. The detailed
comparison with the standard EA replica procedure is performed.
\end{abstract}

\tableofcontents

\section{Introduction}

In the present note we discuss the noncommutative replica approach
to statistical mechanics of disordered systems, proposed in
\cite{kozyrev}. This approach is an alternative to the standard
Edwards--Anderson (or EA)
replica approach introduced in \cite{EA}, see for the review
\cite{SpinGlass}.
Using the operator mean field approach to
noncommutative replica symmetry breaking, we are able to
derive the ultrametric space of states, which describes the
disordered system after the phase transition. In the considered
simplest case the obtained space of states is a $p$--adic disk.

For the review of applications of ultrametricity in physics see
\cite{SpinGlass1}. For introduction to $p$--adic analysis see
\cite{VVZ}, \cite{BS}. $p$--Adic analysis and $p$--adic
mathematical physics attract great interest, see
\cite{VVZ}--\cite{KhrenAlb}.
$p$--Adic analysis
was applied to investigate the spontaneous breaking of the replica
symmetry, cf. \cite{ABK}, \cite{PaSu}.

We investigate disordered models with quenched disorder such as
the Sherrington--Kirkpatrick (or SK) model. We discuss the general
approach to quenched disorder and suggest, that the procedure to
describe quenched disorder, which we call the general (not necessarily EA)
replica procedure, is essentially non unique.

We propose the general definition of replica procedure as a
morphism of noncommutative probability spaces, or the map of
probability spaces, which (in the high temperature limit) will
conserve the correlation functions.

In the standard replica approach quenched disorder is described by the EA
replica procedure \cite{EA}, \cite{SpinGlass}. We construct the
family of examples of replica procedures, all of which are different from
the EA replica procedure. We show, using the Wigner theorem,
see \cite{Wig}, \cite{Voi92}, that in the free case (in the high
temperature limit) the disordered system will be described by the
Fock state over the quantum Boltzmann algebra. For interacting
system (for the finite temperature) the state will become non
Fock.

The quantum Boltzmann algebra arises in the limit of large
stochastic matrices \cite{Wig}, was used in the free probability
\cite{Voi92} and describes the quantum system in the stochastic
approximation \cite{AccLu}--\cite{entangled}.

To describe phase transitions in disordered systems we use
the operator mean field approach and the
free coherent states. The free coherent states were introduced and
investigated in \cite{coherent1}--\cite{coherent3}. In
\cite{represent} corresponding representation of the Cuntz algebra
was constructed.

In the present paper we put the detailed comparison of the
EA and noncommutative replica procedures.
The structure of the paper is as follows.

In Section 2 we remind the EA replica procedure.

In Section 3 we discuss the noncommutative replica procedure of
\cite{kozyrev} and compare it with the EA replica procedure.

In Section 4 we discuss the relation between the high temperature
limit of the noncommutative replica procedure and the Wigner
theorem and propose a general definition of the replica procedure.

In Section 5 we discuss the noncommutative replica symmetry breaking.

In Section 6 we give the conclusion of our results.

\section{The EA replica procedure}

Consider the disordered model with the Hamiltonian $H[\sigma,J]$,
where $\sigma$ enumerates the states of the system and
$J=(J_{ij})$ is the random matrix of interactions with the matrix
elements which are independent Gaussian random variables with the
probability distribution \be\label{J} P[J]=\prod_{i\le j}^N
\exp\left(-{J^2_{ij}\over 2}\right) \ee For simplicity we neglect
here the normalization of the random matrix in the exponent,
taking into account that we may easily reproduce the normalization
by proper rescaling of the correlators.

The averaging over the random interaction $J_{ij}$ is called in
the literature the assumption of self--averaging: to calculate the
expectation value of the observable first we have to take the
average over the system degrees of freedom and then take the
average over the Gaussian stochastic variables $J_{ij}$.

The typical disordered model is the Sherrington--Kirkpatrick (or
the SK) model with the Hamiltonian \be\label{SK}
H[\sigma,J]=-{1\over 2}\,\sum_{i<j} J_{ij}\sigma^i \sigma^j \ee
where the summation runs over the spins $\sigma_{j}$ in the
$d$--dimensional lattice taking values $\pm 1$. Here
$\sigma=\{\sigma_i\}$ describes the orientations of all the spins.

The replica procedure is used to describe disordered models
with quenched disorder. The disorder is called quenched when the Hamiltonian
$H[\sigma,J]$ of the model depends on some typical realization of
the random variable $J$ and from the beginning no averaging on $J$ is assumed.

In this case statistic sum of the disordered system is
$$
Z[J]=\sum_{\{\sigma\}} \exp\left(-\beta  H[\sigma,J]\right)
$$
where $\beta$ is the inverse temperature.

If the averaging of the statistic sum is assumed from the
beginning then the disorder is called annealed. The state for
annealed disordered system is determined by the following annealed
statistic sum \be\label{annealed} \langle\langle
Z[J]\rangle\rangle= \int\sum_{\{\sigma\}} \exp\left(-\beta
H[\sigma,J]\right) \exp\left(-{1\over2}\sum_{i\le j}^N
J^2_{ij}\right) \prod_{i\le j}^N dJ_{ij} \ee In the following we
will omit the $\langle\langle \cdot\rangle\rangle$ brackets.

Remind the EA replica procedure, see \cite{EA},
\cite{SpinGlass}.

The replica procedure \cite{EA} was developed in order to describe
quenched disorder. This procedure consists in introduction of $n$
identical replicas of the original system. This means that we take
the $n$--th degree of the statistic sum
$$
Z^n[J]=\left(\sum_{\{\sigma\}} \exp\left(-\beta
H[\sigma,J]\right)\right)^n
$$
and perform the averaging with the random matrix (\ref{J}).

This implies the following expression for the quenched
statistic sum \be\label{Zn} Z_n=\int\sum_{\{\sigma^a\}}
\exp\left(-\beta \sum_{a=1}^n H[\sigma^a,J]\right) \exp\left(
-{1\over2}\sum_{i\le j}^N J^2_{ij}\right) \prod_{i\le j}^N dJ_{ij}
\ee where we sum over the replicas $\sigma^a$ of the system.

Then one calculates thermodynamical characteristics of system, for
example the free energy, and takes the limit $n\to 0$.

The averaging on $J$ in models with quenched disorder is the
physical assumption of selfaveraging: we assume that the behavior
of the model depending on the typical realization of the disorder
is the same as for the model with averaged statistic sum.

This procedure may be discussed as follows. Consider the
self--averaged free energy
$$
F=-\beta^{-1} \langle\langle \ln  Z\rangle\rangle
$$
Since
$$
\ln x=\lim_{n\to 0}{1\over n}\ln x^n=\lim_{n\to 0}{x^n-1\over n}
$$
the replica procedure implies the following expression for the
free energy
\be\label{Fn}
F_n=-\beta^{-1}{1\over n}  [Z_n-1] =-\beta^{-1}{1\over
n}\left[\int\sum_{\{\sigma^a\}} \exp\left(-\beta \sum_{a=1}^n
H[\sigma^a,J]\right) \exp\left( -{1\over2}\sum_{i\le j}^N
J^2_{ij}\right) \prod_{i\le j}^N dJ_{ij}-1\right]
\ee
Then the quenched free energy will be equal
\be\label{F}
F=\lim_{n\to 0} F_n
\ee

Let us note that (\ref{Fn}), (\ref{F}) in the EA replica approach is
the {\it definition} of the free energy of the disordered system with
quenched disorder. Similar definitions should be made for the
correlators.

In the present paper we introduce the alternative replica procedure
(which we call noncommutative), and discuss the generalizations
of the EA replica procedure.

This suggests the following definition.

\bigskip

\noindent{\bf Definition 1}.\qquad {\sl We call replica procedure the
procedure which transforms the state describing the system with
annealed disorder into the state describing the corresponding
system with quenched disorder.
This procedure does not necessarily coincide with the EA replica procedure.}

\bigskip

The state of a disordered system with quenched
disorder is not determined only by a Hamiltonian of the system. To
describe a state with quenched disorder we also have to describe
the replica procedure. In principle, this procedure may be
non unique, and systems with the same Hamiltonian may have
different quenched states.

This feature, which looks strange, may be justified by the remark,
that to describe quenched states we always use the assumption of
selfaveraging, where we average over the realizations of disorder.
This means that in reality we fix not a Hamiltonian, but a class
of Hamiltonians with similar structure of disorder. Different
replica procedures in this approach may be related to different
kinds of incomplete selfaveraging. Of course, different replica
procedures should describe the states with silmilar behavior, at
least in the high temperature limit. We will formulate the
conditions which should be satisfied by replica procedure in
Definition 8 below.

In the present paper we describe the replica procedure,
called the noncommutative replica procedure (or NRP), which is different
from the EA replica procedure.

\section{The noncommutative replica procedure}

Let us discuss the EA replica procedure using the well known interpretation
of the replica procedure
by nonequilibrium states. Let us assume that we have two
temperatures: one temperature $\beta^{-1}$
for the system degrees of freedom (say for the spins $\sigma$
in the SK model), and the second temperature $\beta'^{-1}$
for the disorder (the matrix $J$
of spin interactions in the SK model). Then we have for the quenched
statistic sum
\be\label{2T}
\tilde Z=\int
\exp\left(-\beta' F[J]\right) P[J]dJ=
\int\exp\left({\beta'\over \beta}\ln Z[J]\right) P[J]dJ=
\int Z^n[J] P[J]dJ
\ee
which, in the limit $n\to  0$, gives the EA replica statistic sum.

Here $n={\beta'\over \beta}$, $F[J]$ is the free energy for the disorder $J$
and $Z[J]$ is the corresponding statistic sum.

Then the limit $n\to  0$ corresponds to the regime when the temperature
of the disorder is much higher than the temperature of the system.
The physical meaning of this property is the independence of the
disorder degrees of freedom $J$ from the system degrees of freedom $\sigma$.

Another possibility to introduce the statistic sum with two temperatures
is to put
$$
\tilde Z=\int Z[J] \left(P[J]dJ\right)^p
$$
and take the limit $p\to\infty$ which corresponds to the limit $n\to  0$
in (\ref{2T})

This discussion leads to the noncommutative replica
procedure (or NRP), introduced in \cite{kozyrev}.

The EA replica procedure is the transformation
\be\label{replica} Z=\int\sum_{\{\sigma\}} \exp\left(-\beta
H[\sigma,J]\right) \exp\left(-{1\over2}\sum_{i\le j}^N
J^2_{ij}\right) \prod_{i\le j}^N dJ_{ij}\mapsto
$$
$$
\mapsto  Z_n=\int\left[\sum_{\{\sigma\}} \exp\left(-\beta
H[\sigma,J]\right)\right]^n \exp\left(-{1\over2}\sum_{i\le j}^N
J^2_{ij}\right) \prod_{i\le j}^N dJ_{ij} \ee together with taking
the limit $n\to 0$.

The noncommutative replica procedure is the transformation
\be\label{ncreplica} Z=\int\sum_{\{\sigma\}} \exp\left(-\beta
H[\sigma,J]\right) \exp\left(-{1\over2}\sum_{i\le j}^N
J^2_{ij}\right) \prod_{i\le j}^N dJ_{ij}\mapsto
$$
$$
\mapsto  Z^{(p)}=\int\sum_{\{\sigma\}} \exp\left(-\beta
H[\sigma,J]\right) \left[\exp\left(-{1\over2}\sum_{i\le j}^N
J^2_{ij}\right) \prod_{i\le j}^N dJ_{ij}\right]^p \ee together
with taking the limit $p\to \infty$.

One can see that in this scheme we have the correspondence
$n=p^{-1}$.

The difference between the above two transformations is that in
(\ref{replica}) we replicate the system degrees of freedom, and in
(\ref{ncreplica}) we replicate the disorder.

Formula (\ref{ncreplica}), of course, is just a rough scheme. In
reality we have to introduce more complicated procedure.

First, instead of the $p$--th degree in $Z^{(p)}$ in
(\ref{ncreplica})
$$
\left[\exp\left(-{1\over2}\sum_{i\le j}^N J^2_{ij}\right)
\prod_{i\le j}^N dJ_{ij}\right]^p
$$
we have to consider the product \be\label{distribution}
\prod_{a=0}^{p-1}\exp\left(-{1\over2}\sum_{i\le j}^N
J^{(a)2}_{ij}\right) \prod_{i\le j}^N dJ^{(a)}_{ij} \ee where
$J^{(a)}_{ij}$ are independent Gaussian random variables
distributed as in (\ref{J}). Note that the product on $a$ is taken
for both the exponent and the differentials $dJ^{(a)}_{ij}$.

Second, in $H[\sigma,J]$ in $Z^{(p)}$ we have to make the
following substitution for $J$: \be\label{Delta} \Delta:
J_{ij}\mapsto {1\over\sqrt{p}}\sum_{a=0}^{p-1}J_{ij}^{(a)} \ee
where $J^{(a)}$, as above, are independent copies of the random
matrix $J$. The map $\Delta$ is an example of coproduct used in
the theory of quantum groups.

Note that the transformation (\ref{Delta}), taken together with
the distribution (\ref{distribution}), is an embedding of
probability spaces (it conserves all the correlation functions).

The described procedure implies for (\ref{ncreplica}) the
following \be\label{ncreplica1}
Z^{(p)}=\int\sum_{\{\sigma\}}
\exp\left(-\beta H[\sigma,\Delta J]\right)
\prod_{a=0}^{p-1}\exp\left(-{1\over2}\sum_{i\le j}^N
J^{(a)2}_{ij}\right) \prod_{i\le j}^N dJ^{(a)}_{ij}
\ee This is
the noncommutative replica expression for the quenched statistic sum.
We arrive to the following definition.

\bigskip

\noindent{\bf Definition 2}.\qquad {\sl The noncommutative replica
expression for correlation functions of quenched disordered system
is given by the correlations of the order parameter $\Delta J$
given by (\ref{Delta}), defined by the statistic sum
(\ref{ncreplica1}), after the limit $p\to\infty$.}

\bigskip

The noncommutative replica procedure and the EA replica
procedure are different examples of replica procedure. In the next section we
discuss properties of the noncommutative replica procedure,
give a general definition of replica procedure and construct
a family of examples.

\section{Large random matrices and the Wigner theorem}

Discuss the noncommutative replica procedure and compare it with
the described below generalization of the Wigner theorem.

The order parameter for the SK model in the NRP approach
is given by the matrix $J=(J_{ij})$,
with matrix elements corresponding to correlation functions
$$
\langle\sigma_i\sigma_j\rangle
$$
After the noncommutative replica procedure this order parameter
is equal to $\Delta J$.

The quenched NRP state with the statistic sum (\ref{ncreplica1})
defines the state on the algebra generated by $p$ matrices with
independent components. In the high temperature regime we have
$\beta\to 0$, and one can neglect the $\beta H$ term in
(\ref{ncreplica1}). In this case the NRP state reduces to the
state described by a variant of the Wigner theorem.

The following result was presented in \cite{ALV}, \cite{Voi92}
and was used in the $N\to\infty$ limit of the matrix model, see
\cite{AV},\cite{ALV}.

Consider the space $P(N,{\bf R})$ of polynomials of symmetric
$N\times N$ matrices over real numbers.
Introduce $p$ independent copies of the space $P(N,{\bf R})$ and
the tensor degree $P(N,{\bf R})^{\otimes p}$. Introduce the state
on $P(N,{\bf R})^{\otimes p}$ in the following way
\be\label{expstate} \langle
f\left(J^{(0)},\dots,J^{(p-1)}\right)\rangle_N=$$ $$={1\over
Z_N}\int \hbox{ tr
}f\left({J^{(0)}\over{N}},\dots,{J^{(p-1)}\over{N}}\right)
e^{-{1\over 2}\sum_{k=0}^{p-1}\hbox{tr }
J_k^2}\prod_{k=0}^{p-1}\prod_{i\leq j}dJ^{(k)}_{ij}; \ee
$$
Z_N=\langle 1\rangle_N.
$$

\bigskip

\noindent{\bf Theorem 3}.\qquad{\sl The limits (\ref{expstate})
exist for each polynomial $f$ and are equal to
$$
\lim_{N\to\infty}\langle
f\left(J^{(0)},\dots,J^{(p-1)}\right)\rangle_N=
(\Omega,f(Q_0,\dots,Q_{p-1})\Omega);
$$
where
$$
Q_a=A_a+A_a^{\dag};
$$
Here $A_a^{\dag}$ and $A_a$ are the quantum Boltzmann
creators and annihilators
and $\Omega$ is the vacuum in the free Fock space:
$$
A_a\Omega=0.
$$

}

\bigskip

The quantum Boltzmann algebra is generated by the Boltzmannian
creation and annihilation operators $A_a $, $A_{a}^{ \dag}$,
$a=0,\dots,p-1$ with the relations \be\label{qB} A_a A_{b}^{
\dag}=\delta_{ab} \ee The Fock representation is constructed in
the free, or quantum Boltzmann, Fock space, generated from the
vacuum $\Omega$, $A_a\Omega=0$, by the action of creators
$A_{a}^{\dag}$.

\bigskip

We see, that the
thermodynamic $N\to\infty$ limit of random matrices
with the distribution (\ref{J}), by the Wigner
theorem \cite{Wig}, \cite{Voi92}, gives rise to the quantum
Bolzmann algebra in the Fock representation.

The order parameter of the noncommutative replica approach,
which is equal to the $\Delta$ of large random matrix,
takes the form of the following operator
from the Fock representation of the quantum Boltzmann algebra
\be\label{Q} Q=\lim_{N\to\infty}{1\over N}\Delta
J={1\over\sqrt{p}}\sum_{a=0}^{p-1}Q_a=
{1\over\sqrt{p}}\sum_{a=0}^{p-1}\left(A_{a}+A_{a}^{\dag}\right)
\ee
where $A_{a}$ and $A_{a}^{\dag}$ are quantum Boltzmann annihilator
and creator correspondingly.

The convergence is understood in the sense of correlators, see
\cite{Wig}, \cite{Voi92}, \cite{ALV}.

In the Fock representation the expectation of this order
parameter is zero:
$$
\langle\Omega,Q\Omega\rangle=0
$$
This state will describe disordered system in the high temperature
regime, when we can neglect the interaction of the system degrees
of freedom. We see that we can reformulate Theorem 3 as follows:

\bigskip

\noindent{\bf Proposition 4}.\qquad {\sl The free theory (the high
temperature regime) of disordered systems is described by the Fock
state over the quantum Boltzmann algebra. For finite temperature
the state will become non Fock. }

\bigskip

Discuss now the noncommutative replica procedure.
Consider the linear coproduct map, which was used in the NRP:
$$
\Delta: P(N,{\bf R})\longrightarrow P(N,{\bf R})^{\otimes p};
$$
\be\label{onX}
\Delta: J\mapsto {1\over\sqrt{p}}\biggl(J\otimes 1\otimes
1\otimes\dots\otimes 1+1\otimes J\otimes 1\otimes\dots\otimes
1+\dots+ $$ $$+1\otimes 1\otimes \dots\otimes 1\otimes J\biggr);
\ee or equivalently
$$
\Delta: J\mapsto {1\over\sqrt{p}}\sum_{a=0}^{p-1} J_a;
$$
where $J_a$ belongs to the $a$--th component of the tensor
product. This map coincides with the map used in the central limit
theorem. For example, for the gaussian random variables $J$ the
map (\ref{onX}) is an embedding of probability spaces (all the
correlation functions are invariant). For large random matrices,
in the thermodynamic (large $N$) limit, the central limit theorem
becomes the free central limit theorem, see \cite{Voi92}. For
instance, the Wigner state will be invariant under (\ref{onX}). We
formulate the following:

\bigskip

\noindent{\bf Theorem 5}.\qquad{\sl The map (\ref{onX}) with the
state  (\ref{expstate}), where we put
$$
f\left(J^{(0)},\dots,J^{(p-1)}\right)= f\left(\Delta J\right);
$$
is an embedding of algebraic probability spaces. In the limit
$N\to\infty$ this embedding becomes the $*$--homomorphism of the
quantum Boltzmann algebra with one degree of freedom maps into the
quantum Boltzmann algebra with $p$ degrees of freedom defined
according to the formula \be\label{algebra}
A\mapsto{1\over\sqrt{p}}\sum_{a=0}^{p-1} A_a,\qquad
A^{\dag}\mapsto{1\over\sqrt{p}}\sum_{a=0}^{p-1} A^{\dag}_a; \ee
and the Fock state $\langle\Omega,x\Omega\rangle$ will map onto
the Fock state $\langle\Omega,x\Omega\rangle_p$ in the quantum
Boltzmann Fock space with $p$ degrees of freedom \be\label{state}
\langle\Omega,x\Omega\rangle\mapsto
\langle\Omega,x\Omega\rangle_p. \ee For instance the map
(\ref{algebra}), (\ref{state}) conserves all the correlators. }

\bigskip

Let us remind the following well known definition.

\bigskip

\noindent{\bf Definition 6}.\qquad {\sl Noncommutative (or
quantum) probability space is a pair $({\cal A},\phi)$, where
${\cal A}$ is an algebra over the field of complex numbers with
unit and involution, and $\phi$ is a positive normed state on the
algebra ${\cal A}$:
$$ \phi(a^*a)\ge 0, a \in {\cal A},\qquad \phi(1)=1$$ }

\bigskip

Category of noncommutative probability spaces is the category,
where the objects are noncommutative probability spaces, and the
morphisms a $*$--homomorphisms of algebras, conserving correlation
functions.

This means that if $f$ is a morphism of noncommutative probability
spaces
$$
f:({\cal A},\phi)\mapsto ({\cal B},\psi)
$$
and $P\in {\cal A}$ is a (noncommutative) polynomial
$P(a_1,\dots,a_k)$ in ${\cal A}$, then the properties of a
homomorphism
$$
f(P(a_1,\dots,a_k))=P(f(a_1),\dots,f(a_k))
$$
involution
$$
f(a)^*=f(a^*)
$$
and conservation of correlators
$$
\psi(f(a))=\phi(a)
$$
are satisfied.

We call the morphism $f$ an embedding, if it is an embedding as a
map of algebras.

\bigskip

The Wigner theorem and related results (theorems 3, 5, proposition
4 ) describe the free case (or the high temperature limit) of
disordered systems. The NRP in this case takes the form
(\ref{onX}) (or (\ref{algebra}) in the $N\to\infty$ limit), where
for the states the NRP takes the form (\ref{state}) (in the
$N\to\infty$ limit).

Let us discuss now the properties of the NRP in the interacting
case. In this case the NRP for observables has the same form,
while for the states the NRP state will be determined by the NRP
statistic sum (\ref{ncreplica1}).

If in (\ref{ncreplica1}) one can interchange the summation over
$\sigma$ and the integration over $J_{ij}$ we get for the
expectation of $(\Delta J)^k$:
$$
\langle (\Delta J)^k\rangle={1\over Z^{(p)}}\sum_{\{\sigma\}} \int
(\Delta J)^k \exp\left(-\beta H[\sigma,\Delta J]\right)
\prod_{a=0}^{p-1}\exp\left(-{1\over2}\sum_{i\le j}^N
J^{(a)2}_{ij}\right) \prod_{i\le j}^N dJ^{(a)}_{ij}
$$
Then, expanding $\exp\left(-\beta H[\sigma,\Delta J]\right)$ into
series over the degrees of $\beta$ and interchanging again (if
possible) the summation with integration, we get that the NRP
(\ref{Delta}) will conserve all the expectations of the degrees of
the random matrix $J$:
$$
\langle (\Delta J)^k\rangle=\langle  J^k\rangle
$$
where at he LHS of this formula the state is defined by the NRP
statistic sum (\ref{ncreplica1}), and at the RHS the state is
defined by the annealed statistic sum (\ref{annealed}).

The analyticity of the statistic sum with respect to $\beta$ and
the possibility to interchange summations with integration in the
formula above are related to the existence of phase transitions in
the system: if there is no phase transitions, then the statistic
sum is analytic with respect to $\beta$. In this case the
arguments above are correct and all the correlation functions are
invariant under the NRP. If we have phase transitions then the NRP
will modify the state.

Let us formulate now the following proposition.

\bigskip

\noindent{\bf Proposition 7}.\qquad {\sl For the case when there
is no phase transition the NRP defined by (\ref{Delta}),
(\ref{ncreplica1}) defines the state equivalent to the annealed
state. After phase transition the state defined by the NRP will be
not equivalent to the annealed state.}

\bigskip

Let us note that the EA replica procedure has similar properties:
it conserves the correlators in the high temperature limit.

Summing up, we see that the Wigner theorem allows to compute the
thermodynamic limit of the SK model in the high temperature
regime. We introduce the noncommutative replica transformation
(\ref{onX}) and show, that in the high temperature regime it
realizes an equivalent description of the disordered model (by
Theorems 3 and 5). The Proposition 7 shows, that this will be also
true for finite temperatures above the temperature of phase
transition.

One can easily construct a family of generalizations of
transformation (\ref{onX}) which are also the embeddings of
probability spaces.

For instance, one can consider the map \be\label{onX1} \Delta':
J\mapsto {1\over\sqrt{p}}\sum_{a=0}^{p-1} c_a J_a; \ee where $c_a$
are complex valued coefficients, which should satisfy the
condition
$$
\sum_{a=0}^{p-1} |c_a|^2 =p
$$
which guarantee that the map (\ref{onX1}) with the Wigner state
used in Theorem 5 is an embedding of probability spaces).

To make this map the replica procedure for the disordered system,
one has to extend it on the states, by defining the quenched
statistic sum \be\label{que1} Z'^{(p)}=\int\sum_{\{\sigma\}}
\exp\left(-\beta H[\sigma,\Delta' J]\right)
\prod_{a=0}^{p-1}\exp\left(-{1\over2}\sum_{i\le j}^N
J^{(a)2}_{ij}\right) \prod_{i\le j}^N dJ^{(a)}_{ij} \ee Different
replica procedures of the form (\ref{onX1}), (\ref{que1}) should
correspond to physically different disordered systems, which have
different behavior after phase transition.

Let us discuss the obtained results from the point of view of
nocommutative probability theory. In the theory of disordered
systems we the following probability space. The algebra is the
$*$--algebra ${\cal J}$ generated by large symmetric random
matrix, and the state $\langle\cdot\rangle$ (depending on the
inverse temperature$\beta$) is the annealed Gibbs state generated
by the annealed statistic sum (\ref{annealed}). Therefore the
annealed case is described by a commutative probability space. We
will see that the quenched state should be described by
noncommutative probability space.

The discussed in the present paper replica procedures are the maps
which, in the high temperature regime, conserve the correlation
functions. This suggests to formulate the following definition.

\bigskip

\noindent{\bf Definition 8}.\qquad {\sl We call the replica
procedure the family $f_{\beta}$ of maps of the probability space
$({\cal J},\langle\cdot\rangle)$, parameterized by the inverse
temperature $\beta$, into the other probability space $({\cal
A},\phi)$ (which also depends on $\beta$), which in the high
temperature regime $\beta\to 0$ becomes a morphism in the category
of quantum probability spaces. }

\bigskip

This definition means, that for high temperatures any quenched
states, generated by the replica procedure, should be equivalent
to the annealed state. For low temperature this equivalence will
broken. For instance, different replica procedures may generate
different quenched states. This may be discussed as the
consequence of different self--averaging of the quenched disorder.

The noncommutative replica procedure, discussed in the present
paper, maps the annealed probability space $({\cal
J},\langle\cdot\rangle)$ into the quenched, or replica probability
space $({\cal J}_{\infty},\langle\cdot\rangle)$, generated by
infinite number of random matrices $J^{(a)}$, $a=0,\dots$ with
independent matrix elements and the state generated by the replica
statistic sum (\ref{ncreplica1}). In the limit of large random
matrices this probability space becomes the quantum Boltzmann
algebra with infinite number of degrees of freedom, where the
state is the $N\to\infty$ limit of the replica state
(\ref{ncreplica1}).

Since two different replica procedures $f_{\beta}$ and $g_{\beta}$
(morphisms of probability spaces)  may have noncommuting images,
we see that the general approach in the replica theory should be
described in the framework of noncommutative probability theory.

\section{The noncommutative RSB: the operator mean field theory}

To describe phase transitions in disordered systems in the EA replica
approach the procedure of the replica symmetry breaking, or RSB, is used
\cite{SpinGlass}. In this procedure the properties of the system
are described by the Parisi replica matrix, which in $n\to 0$
limit becomes the operator in infinite dimensional space.
$p$--Adic parameterization for the Parisi matrix was obtained
by Avetisov, Bikulov, Kozyrev \cite{ABK},
and by Parisi, Sourlas \cite{PaSu}.

In the present section we discuss an approach to describe phase transitions
in disordered systems, called the noncommutative replica symmetry breaking,
introduced in \cite{kozyrev}.
This approach is a kind of a mean field approach, where the mean field
(the order parameter) is an operator. The operator mean field approach
was used, for instance, in the Bogoliubov approach in theory
of superconductivity and superfluidity \cite{super}.
In this approach the phase transition was described by the quantum
mean field condition
$$
A\Psi=\lambda\Psi
$$
where $A$ is the Bose annihilation operator.
This condition means that the Bose condensate is in the coherent state.

In the theory of superconductivity and superfluidity the operator
mean field theory \cite{super} describes the changing of the
kinematics of the system, which results in the effects of
superconductivity and superfluidity.

For quenched disordered systems the statistics is described by
the replica transformed large random matrices. By the Wigner theorem,
large random matrix has the quantum Boltzmann statistics, and after
the noncommutative replica transformation the order parameter becomes
$$
Q={1\over{\sqrt{p}}}\sum_{a=0}^{p-1} \left(A_{a}+A_{a}^{\dag}\right),\qquad
A_{a}A_{b}^{\dag}=\delta_{ab}
$$
The condition of the noncommutative mean field, similar to used in
\cite{super}, takes the form
\be\label{mean}
{1\over{\sqrt{p}}}\sum_{a=0}^{p-1} A_{a} \Psi=\lambda\Psi
\ee
We will see, that condition (\ref{mean}), due to the quantum Boltzmann
statistics of the order parameter for disordered systems,
will lead to nontrivial effects already on the level of statics.

Let us describe now the noncommutative replica symmetry breaking procedure.
It includes the following constructions.

\bigskip

\noindent{\bf The first step}\qquad To describe the phase
transition, one has to consider the representation in which the
order parameter has non zero expectation. To do this, we apply
the noncommutative mean feld approach and consider
\cite{kozyrev} the free coherent states, constructed in
\cite{coherent1}--\cite{coherent3}. The free coherent state $\Psi$
is the eigenvector of the free annihilation \be\label{coherent}
A\Psi=\lambda\Psi, \qquad
A={1\over\sqrt{p}}\sum_{a=0}^{p-1}A_{a},\quad \lambda \in {\bf R}
\ee Note that here $\Psi$ is a function of $\lambda$.

One easily observes that the expectation of the order parameter
$Q$ between the free coherent states is non--zero:
$$\langle Q\rangle\ne 0$$

In the works
\cite{coherent1}--\cite{coherent3} it was shown that, in the
renormalized scalar product,
$$
(\Psi,\Phi)=\lim_{\lambda\to 1-0}\langle\Psi,\Phi\rangle
$$
the space of the free coherent states is isomorphic to the space
of (complex valued) generalized functions on $p$--adic disk $Z_p$.

This shows that the thermodynamic limit $N\to \infty$, together
with the procedure of noncommutative replica symmetry breaking
$\langle Q\rangle\ne 0$ in the noncommutative mean field approach,
implies the derivation of the ultrametric (and even $p$--adic) space of states,
which was postulated in the EA replica approach in the procedure
of the replica symmetry breaking. All the procedure of
construction of the ultrametric space in the noncommutative
replica approach is encoded into the simple algebraic condition
(\ref{coherent}). No conditions of the kind $n\to 0$ or
$p\to\infty$ are relevant to ultrametricity.

In \cite{kozyrev}, \cite{coherent3} the condition (\ref{coherent})
was discussed as the equation of the noncommutative line
\be\label{qline}
A=1, \qquad A={1\over\sqrt{p}}\sum_{a=0}^{p-1}A_a
\ee
in the noncommutative plane with the coordinates $A_a$.
The results of \cite{coherent1}--\cite{coherent3} in this language
may be reformulated as the equivalence of the noncommutative
line (\ref{qline}) and $p$--adic disk.

This, together
with the arising of the noncommutative quantum Boltzmann algebra
in the Wigner theorem, explains the name noncommutative replica
approach.

\bigskip

\noindent{\bf The second step}\qquad We showed already that the
degrees of freedom of the system in the $N\to\infty$ limit will be
described by the quantum Boltzmann algebra (\ref{qB}), which
before the phase transition acts in the Fock representation, and
after the phase transition will act in the non Fock
representation, related to the free coherent states. In
\cite{kozyrev}, \cite{represent} the following expressions for
this representation were obtained: the operators from the quantum
Boltzmann algebra
$$
A_aA_b^{\dag}=\delta_{ab}.
$$
in this representation satisfy the Cuntz relation
$$
\sum_{a=0}^{p-1}A_a^{\dag} A_a=1.
$$
and are realized in the space $L^2(Z_p)$ of quadratically
integrable functions on $p$--adic disk, where act as follows:
\be\label{Adagpadic} A^{\dag}_a \xi(x)=\sqrt{p}\theta_1(x-a)
\xi([\frac{1}{p}x]); \ee \be\label{Apadic} A_a
\xi(x)=\frac{1}{\sqrt{p}}\xi(a+px). \ee Here
$$
[x]=x-x(\hbox{mod }1)
$$
for $x\in Q_p$ is the integer part of $x$. $\theta_1(x-a)$ is the
characteristic function of the $p$--adic disk with the center in
$a$ and the radius $p^{-1}$.

Physically this representation will describe the disordered system
in the limit of zero temperature.

\section{Discussion of the NRP}

Let us make the conclusion of our discussion. In the present note
we described the noncommutative replica procedure and compared it
with the EA replica procedure.

We conjecture, that the transition from annealed to quenched
disorder can be described by the noncommutative replica procedure
(\ref{Delta}), (\ref{ncreplica1}).

For the high temperatures (above the temperature of phase
transition) the NRP is an embedding of probability spaces, that
means that (\ref{Delta}) conserves the correlation functions:
\be\label{XDX} \langle f(J)\rangle=\langle f(\Delta J)\rangle \ee
Here $f(J)$ lies in the algebra generated by the large random
matrix $J$, the state at the LHS is given by (\ref{J}), and the
state at the RHS is given by (\ref{distribution}).

We will say that in (\ref{XDX}) the observable $J$ at the LHS lies
in the annealed algebra of observables, and $\Delta J$ at the RHS
lies in the quenched algebra (and correspondingly for the states).

The noncommutative replica procedure implies that in the limit of
infinite temperature, by the Wigner theorem, the state of a
disordered system will be equal to the vacuum state in the quantum
Boltzmann Fock space.

We propose the definition of the replica procedure as the map
between noncommutative probability spaces which becomes a morphism
for high temperatures.

We described the noncommutative replica symmetry breaking
condition in the operator mean field approach
$\langle Q\rangle\ne 0$, $Q=A+A^{\dag}$,
$A={1\over\sqrt{p}}\sum_{a=0}^{p-1} A_{a}$ as the equation of quantum line
$A\Psi=\lambda\Psi$. The application of the theorem about
the isomorphism between the quantum line and $p$--adic disk
allows to derive the ultrametric
space of states which is postulated in the Parisi replica
symmetry breaking approach \cite{SpinGlass}.

The correlation functions of the disordered model after the
noncommutative replica symmetry breaking will be calculated using
the $p$--adic representation of the Cuntz algebra
(\ref{Adagpadic})--(\ref{Apadic}) (for zero temperature).

\bigskip

\centerline{\bf Acknowledgements}

The author would like to thank I.V.Volovich, V.A.Avetisov,
A.H.Bikulov and A.Yu.Kh\-ren\-ni\-kov for discussions and valuable
comments. This work has been partly supported by INTAS YSF 2002--160 F2,
CRDF (grant UM1--2421--KV--02), and The Russian Foundation for
Basic Research (projects 02--01--01084 and 00--15--97392).

\end{document}